\documentclass[useAMS,usenatbib,usegraphicx]{mn2e}
\usepackage{times}
\title[The near-infrared spectrum of Mrk~1239]{The near-infrared spectrum of Mrk~1239: direct evidence of the dusty torus?}
\author[A. Rodr\'{\i}guez-Ardila and X. Mazzalay]{A. Rodr\'{\i}guez-Ardila$^{1}$\thanks{Visiting Astronomer at the Infrared Telescope Facility, which is operated by the University of Hawaii under Cooperative Agreement no. NCC 5-538 with the National Aeronautics and Space Administration, Office of Space Science, Planetary Astronomy Program.} and X. Mazzalay$^{2}$\\
$^{1}$Laborat\'orio Nacional de Astrof\'{\i}sica/MCT, Rua dos Estados Unidos 154, CEP 37500-000, Itajub\'a, MG, Brazil. E-mail:aardila@lna.br\\
$^{2}$IATE, Observatorio Astron\'omico C\'ordoba, Laprida 854, X5000BGR, 
C\'ordoba, Argentina. E-mail:ximena@oac.uncor.edu\\
}
\begin{document}

\date{}

\pagerange{\pageref{firstpage}--\pageref{lastpage}} \pubyear{Accepted to MNRAS Letters, 2006}

\maketitle

\label{firstpage}

\begin{abstract}
We report 0.8$-$4.5~$\mu$m SpeX spectroscopy of the 
narrow-line Seyfert 1 galaxy Mrk~1239. The spectrum is outstanding 
because the nuclear continuum emission in the near-infrared is
dominated by a strong bump of emission peaking at 2.2~$\mu$m,
with a strength not reported before in an AGN.
A comparison of the Mrk~1239 spectrum to that of Ark~564 allowed 
us to conclude that the continuum is strongly reddened by E(B-V)=0.54. The excess 
of emission, confirmed by aperture photometry and additional NIR 
spectroscopy, follows a simple blackbody curve at T$\sim$1200~K.
This suggest that we may be observing direct evidence of dust heated near 
to the sublimation temperature, likely produced by the 
putative torus of the unification model. Although other alternatives are
also plausible, the lack of star formation, the strong polarization and
low extinction derived for the emission lines support the scenario 
where the hot dust is located between the narrow line region and the 
broad line region.
\end{abstract}


\section{Introduction}

Unified schemes of active galactic nuclei (AGN) invoke the presence
of an obscuring dusty torus around the central engine, giving rise to
type~1 objects for for pole-on viewing and type~2 objects in
edges-on sources. This obscuring structure would also absorbs a
significant fraction of the optical/UV/X-ray continuum of the central
source and should radiate back this energy at IR wavelengths. In 
fact, dust reprocessing is regarded as the most likely source of the 
strong near- and mid-infrared (NIR and MIR, 
respectively) continuum emission in radio-quiet 
quasars and Seyfert galaxies. Observational evidence favoring models 
in which the IR continuum between 1 and 10~$\mu$m is predominantly or 
entirely dominated by emission from heated dust is abundant 
\citep{em86,ba87,als03}. Indeed, evidence of the presence
of hot dust near the sublimation temperature in Seyfert~1 galaxies
comes from the observation of a peak of emission with central wavelength between 
2.2$-$3.5~$\mu$m found from JHKL and L' photometry  \citep{gla92,ma98,ma00} 
in a few objects. Moreover, very recently, \citet{ro05} found in the inner 250~pc of 
Mrk~766, a narrow-line Seyfert~1 galaxy,  this same excess of emission
by means of NIR spectroscopy. They were able to determine accurately the 
form of the NIR bump, confirming that it follows a simple blackbody function at 
T=1200~K. They found that the host dust emission 
accounted for up to 28\% of the total NIR continuum flux in that object.

From above, the growing observational evidence of
NIR thermal emission at temperatures close to the sublimation 
temperature of graphite grains in Seyfert~1 galaxies strongly
supports the unified models for AGNs. Meanwhile, additional 
evidence is needed and many open questions have to be answered.  
In particular, it is necessary to investigate if the thermal
emission could be related to a compact dust/molecular thick 
torus like the ones in the unified models of \citet{pk92a} and \citet{ef95}, for instance,
or if it results from emission by hot dust (T$>$900~K) mixed with gas
in the NLR/BLR interface region, shielded from the intense UV 
radiation field \citep{ma00}.

Here, we contribute to this discussion by presenting 
the most outstanding evidence of a NIR bump reported to date. 
It corresponds to the one displayed by Mrk~1239, a compact galaxy,
classified as narrow-line Seyfert 1 (NLS1) by \citet{op85}. 
Data collected over the past years point out
that Mrk~1239 is indeed dusty. \citet{good89}, for instance,
reports that it is one of the three galaxies that 
display the largest percentage of polarization, both in the line 
and continuum, in the sample of 18 NLS1 he studied. \citet{sm04}
modeled the polarization nature of this object and found that 
it was one of the rare cases of Seyfert~1 galaxies that appear
to be dominated by scattering in an extended region along the 
poles of the torus. According to their results, the line-of-sight
to the nucleus would pass through the relatively tenous upper 
layers of the torus, extinguishing the continuum and BLR
emission. This radiation, would be polarized further off by
the dust located above or below the torus.

X-rays observations also confirm the dusty nature of Mrk~1239. XMM-Newton EPIC 
PN data suggest two light paths between the continuum source and the 
observer, one indirect scattered one, which is less absorbed, and a 
highly absorbed direct light path, in agreement to the 
wavelength-dependent degree of polarization in the optical/UV. 
Moreover, Mrk~1239 is classified as a 
60~$\mu$m peaker \citep{hdr99} because of
its ``warm'' far-infrared colours and spectral energy distribution
peaking near 60~$\mu$m. They attributed these properties to 
dust-obscured active galactic nuclei \citep{ke94,he95}, 
the obscuring material likely
associated to the putative torus of the unified model.

This letter is organized as follows. In Section~\ref{obs} we 
describe the observations, data reduction and resulting spectrum. Section~\ref{red}
determines the internal reddening affecting
the nuclear spectrum of Mrk~1239 and compares the observed 
NIR SED with that of Ark~564. It also analyzes the
different components that contribute to the observed
continuum. Section~\ref{dust} examines the hot dust
hypothesis for Mrk~1239 in the light of the strong thermal
NIR excess of emission detected. Conclusions are in Section~\ref{final}. 
Throughout this work we adopt a Hubble constant of 
H$_o$=75~km\,s$^{-1}$~Mpc$^{-1}$.

\section{Observations, data reduction and results} \label{obs}

\begin{figure}
\includegraphics[width=86mm]{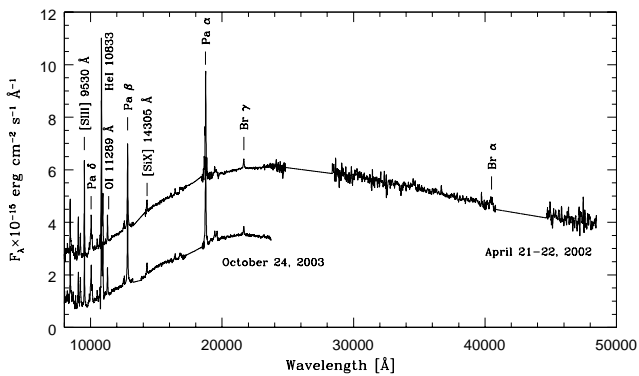}
\caption {Observed NIR spectra of Mrk~1239 in the rest-frame of the
galaxy. The spectrum taken on October 24 has been offset by 2$\times$10$^{-15}$
erg~cm$^{-2}$~s$^{-1}$~\AA$^{-1}$ relative to that of April 21 for 
visualization purposes. The most important permitted and forbidden lines 
are marked. Note the strong excess of emission present in both spectra,
starting at 10000~\AA, with peak at 22000~\AA.
}\label{mrk1239nir}
\end{figure}

NIR spectra of Mrk~1239 in the intervals 0.8$-$2.4~$\mu$m and
2.0$-$4.9~$\mu$m (hereafter SXD and LXD, respectively) were 
obtained at the NASA 3\,m Infrared Telescope Facility (IRTF)
with the SpeX spectrograph \citep{ray03}, atop Mauna Kea, on 
April 21, 2002 (UT) and October 24, 2003 (UT) for
the SXD data and on April 22, 2002 (UT) for the LXD spectrum.
The detector consisted
of a 1024$\times$1024 ALADDIN 3 InSb array with a spatial scale of
0.15$\arcsec$/pixel. A 0.8$\arcsec\times$15$\arcsec$ slit, 
oriented east-west, was used during the observations. The
spectral resolution was 360~km\,s$^{-1}$ at both setups.
The total on-source exposure times amount to 1900\,s and 1200\,s for the April
21 and 22, 2002, and 1200\,s for the SXD observation of October 24, 2003. 
The signal within the central 
0.8$\arcsec\times$1$\arcsec$ (1$\arcsec$=380~pc) was summed up to obtain the nuclear 
spectrum. The light distribution of the galaxy was found to be cuspy, 
being dominated by the unresolved emission from the AGN.
The data reduction, 
extraction and calibration procedures were done using
the in-house software Spextool \citep{cvr04} and Xtellcor \citep{vcr03}, 
provided by the IRTF
Observatory. The spectra in the 
SXD and LXD settings, observed on consecutive nights, were
merged to form a single 0.8$-$4.9~$\mu$m spectrum. The agreement
in the continuum level in the overlapping region was excellent, 
with less than 5\% of uncertainty.
The SXD spectra obtained in October 2003 also agreed, within 10\% of
uncertainty, with the continuum level measured in the observation
taken in 2002. 

Figure~\ref{mrk1239nir} show the final
NIR spectra of Mrk~1239 in the rest-frame of the object. 
In the regions where the atmospheric transmission drops to zero, 
a straight line was interpolated to connect the adjacent bands.
It can be seen that the NIR spectrum 
is dominated by a strong bump of emission, peaking at 22000~\AA,
also present in the October 24, 2003 observation. Also
prominent in the spectra are emission lines of H\,{\sc i}, He\,{\sc i}
and [S\,{\sc iii}]. No absorption lines that may indicate the
presence of circumnuclear stellar population were detected. 
Since the main focus of this work is the
continuum emission, the analysis of the line spectra
is left for a future publication. To our knowledge, the only 
NIR spectroscopy available in the literature on this source
is the {\it J} and {\it K} spectra of \citet{hdr99}. Because of the 
non-photometric conditions in which they
were taken, there is a lack of continuity in the continuum level
between the bands, preventing us to make any meaningful comparison. 
However, in Fig.~2g of \citet{hdr99}, 
the $J$ continuum rises toward longer wavelengths. In the
$K$-band, it continues rising up to 2.2~$\mu$m, where it seems
to become flat, in accordance to our observations.  

In addition to the NIR data, long-slit optical 
spectroscopy on Mrk~1239 was obtained on the nights of March 19 and
20, 2002, with the Cassegrain spectrograph attached to the 2.15~m 
telescope of the CASLEO Observatory, Argentina. The spectra cover 
the interval 3700--9600~\AA\ and were obtained
with a 2$\arcsec$ slit width and a 300~l/mm grating. The extraction
and reduction process followed the standard IRAF procedure. As in
the NIR region, the light profile distribution of the galaxy 
is dominated by the unresolved emission of the AGN. The aim of the 
optical data is to complement the NIR spectrum to map the 
optical-NIR continuum of this source. No significant variation
(less than 5\%) in the level of the continuum
emission was detected in the overlapping region between the
optical and NIR spectra (0.8$-$0.95~$\mu$m).

\section{The internal extinction in Mrk~1239 and the continuum emission}\label{red}

\begin{figure}
\includegraphics[width=88mm]{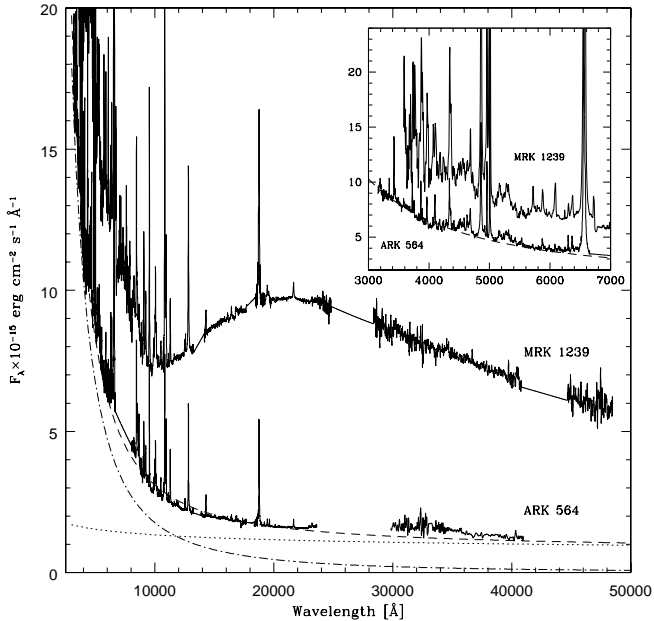}
\caption {Comparison of the optical-NIR spectrum of Mrk~1239, after 
being derreddened by E(B-V)=0.54, with that of Ark~564. The box in the upper 
right corner shows a zoom in the optical region of both galaxies. The 
continuum emission of the latter object was modeled in terms of a sum 
of two power laws (dashed line): one that dominates the optical region 
(dot-dashed line) and another that dominates the NIR (dotted line).
Note the lack of significant NIR excess of emission in Ark~564.
}\label{optnir}
\end{figure}

In order to properly interpret the strong
excess of NIR emission in Mrk~1239, it is first necessary to
de-redden its optical-NIR continuum. To this purpose, we used
as reference, the spectrum of the NLS1 galaxy Ark~564, 
whose continuum emission is well known and we assumed as 
typical of a NLS1 galaxy. Flux-calibrated FOS HST spectra in the
optical region and SpeX IRTF spectra \citep{crv03}
were employed.

We started by derreddening the Ark~564 data by Galactic 
(E(B-V)=0.03) and internal extinction (E(B-V)=0.14), as
determined by \citet{cr02}. 
The reddening law of \citet{ccm89} with $R_{\rm V}$=3.1 was used
to this purpose. Then, assuming that the form and slope of the 
extinction-corrected optical continuum in Ark~564
represents the intrinsic continuum of Mrk~1239, we de-reddened 
the observed spectra of the latter object
(already corrected by a Galactic E(B-V) of 0.065) 
in small steeps until the form of its optical continuum
matches that of former. An E(B-V)=0.54~mag was
necessary and was taken as the intrinsic reddening of Mrk~1239. 
This value is in excellent agreement to the one
predicted by the polar scattering region scenario (E(B-V)$\geq$0.36)
proposed by \citet{sm04} if that mechanism becomes the dominant source 
of polarization.
Note that had we used the NIR H\,{\sc i} line ratios to 
determine the extinction, we would have obtained zero
extinction. The measured B$\gamma$ flux is 2.76$\pm0.20\times10^{-14}$ 
erg~cm$^{-2}$~s$^{-1}$ while that of Pa$\beta$ is 1.91$\pm0.1\times10^{-13}$
erg~cm$^{-2}$~s$^{-1}$. Thus, the ratio B$\gamma$/Pa$\beta$
equals 0.14$\pm$0.01, somewhat lower than the intrinsic Case B
value (0.17). This apparent inconsistency can be explained if
NIR BLR hydrogen lines strongly deviates from Case B due to
collisional and radiation transfer effects, as it the case 
for the optical lines.

Figure~\ref{optnir} shows the extinction corrected spectrum 
of Mrk~1239.
It can be seen that after dereddening, the optical continuum 
of Mrk~1239 has the same steepness as that of Ark~564. Note
the broad bump of emission starting at $\sim1$~$\mu$m, with 
peak at 22000~\AA. It dominates the NIR continuum distribution
in Mrk~1239 but is absent in Ark~564. 

Figure~\ref{optnir} also shows
that at $\sim1$~$\mu$m there is a break in the
steepness of the continuum. If the UV-optical continuum is 
well described by the a power-law of the form 
F$_{\lambda} \propto \lambda^{\alpha}$, the extrapolation of this
function to the NIR region lacks of enough power to explain the
NIR continuum emission. In Ark~564, for instance, where the
strong bump of emission is not observed, an additional power
law is necessary in order to reproduce the NIR SED. If we
consider the continuum of Ark~564 as typical for a NLS1 galaxy, the break 
represents the end of the long wavelength side of the
continuum attributed to the central engine and the
onset of the thermal emission from dust grains. 
The fact that Ark~564 displays a low percentage of
polarization \citep{good89} both in line and 
continuum, indicates that the dust content in the BLR
of this object is small (also confirmed by the small
value of intrinsic extinction). It means that in Ark~564
we are observing the true near-infrared shape of the
big blue bump component, which up to date is essentially unknown
\citep{kab05}.

\subsection{The components of the NIR continuum emission in Mrk~1239}

The continuum emission displayed by Mrk\,1239 in the interval 
0.8$-$4.9~$\mu$m is complex and rather different from the one
displayed by other NLS1 galaxies (see for example, the NIR
spectra of other AGN presented by \citealt{ro02}). A comparison with
the continuum of Ark~564 is, in fact, striking.

In order to characterize the NIR continuum emission in
Mrk\,1239, we used, as reference, the much simpler optical-NIR
continuum of Ark~564. To this purpose, a composite function
described by the sum of two power-laws of the
form F$_{\lambda} \propto \lambda^{\alpha}$ was fitted 
to the data using the task {\it nfit1d} of the STSDAS package of IRAF.
This composite function was chosen because it was clear that
a single power-law cannot reproduce the optical-NIR continuum.
Special care was taken to not include
emission lines in the spectral windows used
in the fit. The results are shown in Figure~\ref{optnir}.
The derived spectral indices are $\alpha_{\rm opt}=-1.93$
and $\alpha_{\rm NIR}=-0.2$ for the optical and
NIR regions, respectively.  From Figure~\ref{optnir}, we see
that this composite function cannot describe the continuum 
emission in the 1$-$4~$\mu$m 
region of Mrk~1239. The strong excess of emission that rises
above the level predicted by the optical plus NIR power-laws
indicates the necessity of an additional 
component.

\begin{figure}
\includegraphics[width=86mm]{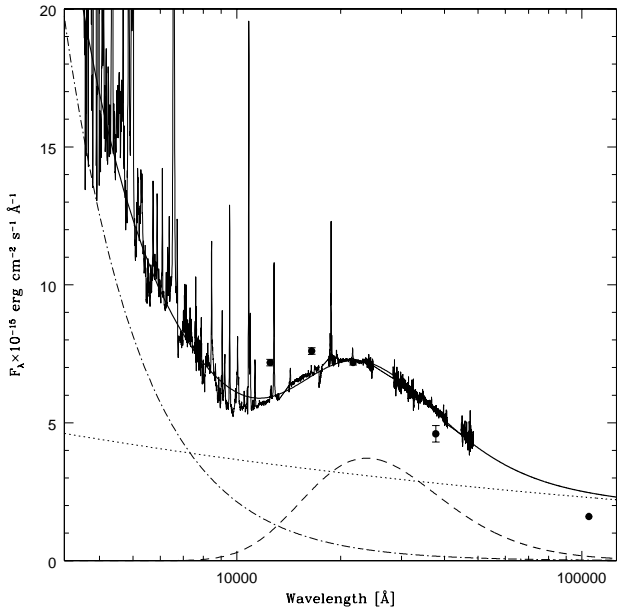}
\caption {Intrinsic optical, NIR and MIR SED of Mrk\,1239.
The full circles are aperture photometry. In the JHK bands, they
were taken from 2MASS. For the L and N region, they were taken 
from the literature (see text). The dotted-dashed and the dotted curves are the
optical and NIR power-law functions, respectively, found for Ark~564. 
The dashed curve is the blackbody distribution of $T_{\rm bb}$=1210~K that
best reproduces the NIR bump. The thick line represent the sum of 
these three components. \label{powl}}
\end{figure}

An inspection to the overall shape of the bump 
suggests that it approaches that of a blackbody distribution. In order
to test this hypothesis, we fitted a composite function $-$two power-laws,
plus a Planck curve, to the observed optical-NIR SED.
Additional constraints to the fit was imposed by adding 
photometric points for the L (3.5~$\mu$m) and N (10.5~$\mu$m)
bands, taken from the literature \citep{spi95,mai95}. 
In the fit, the power-law indices were constrained 
to the values found for Ark~564 (see above) while the temperature and
amplitude of the blackbody function were left as free parameters.
The results, displayed in Figure~\ref{powl}, show that
a composite function, with a blackbody of 
temperature $T_{\rm bb}$=1210~K, provides an excellent description
to the the excess of emission over the power-laws. It should be
noted here that the use of blackbody instead of a greybody is 
preferred because of the smaller number of free parameters 
that the former function requieres. In the absence of observational
constraints that justifies the use of a more complex function, they
choice is for the one that offers an adequate solution with the 
minimum number of parameters to fit. See also Sect.~\ref{dust}

What is the origin of each of the three continuum components, 
all emitted in the inner 380\,pc of the AGN? 
We associate the power-law that dominates the
optical region to the long-wavelength side of the UV/optical
continuum component, often called the big blue bump (BBB),
thought to be emitted from an accretion disk around a supermassive
black hole \citep{mal83,zhe97,con03}.
The NIR power-law is 
likely associated to the blue-end
tail of the much broader infrared excess that dominates
the IR SED of AGN, usually attributed to cold dust 
($T \sim$40-80~K), warmer than dust in normal spiral 
galaxies \citep{em86}, with peak emission
at 60~$\mu$m, and observed in Mrk~1239 \citep{hdr99}.

The third component, well described
by a Planck function of $T_{\rm bb}\sim$1200~K
will be 
examined in the following section. Unfortunately, the 
lack of a spectroscopic survey of AGNs in the NIR prevents 
us to determine if that feature is common in Seyfert galaxies.
The only reports to date of a similar feature comes from spectroscopy 
observations of I\,Zw1 \citep{rudy00}, Mrk\,478 \citep{rudy01} 
and Mrk\,766 \citep{ro05}, notably all NLS1 galaxies. 
In the former two objects, an excess of emission redwards
of 1.3~$\mu$m over the underlying featureless continuum, is observed. 
The I\,Zw1 and Mrk\,478 data, however, are limited to 2.4~$\mu$m.
It is not possible to say what happens beyond that 
point. Note that the narrowness of   
this component turns it difficult to
be detected in broad band photometry studies based on
IRAS observations, for instance. Its presence, however, can be confirmed
by means of 2MASS photometry. In Figure~\ref{powl}, 
we have plotted the {\it JHK} fluxes (full circles) reported 
for Mrk~1239, taken from NED. No photometry is the optical is available. 
Overall, the photometric points follow
the NIR excess. The small overestimation of 
the photometric {\it J} and {\it H} flux over the spectroscopic one ($\sim$16\%) 
can be due to line emission (H\,{\sc i} and He\,{\sc i}) and probably, 
underlying continuum emission from the host 
galaxy. Recall that the 2MASS data plotted is extracted from 
aperture photometry, of radius of 14$\arcsec$, 
which may include contribution from the host-galaxy.

\section{The hot dust hypothesis}\label{dust}

The adequate representation of the NIR bump
by a blackbody distribution leads us to propose
that it is due to emission from hot dust grains. The temperature derived
from the fit, $T_{\rm bb}$=1200~K, is close to
the evaporation temperature of graphite grains, T$\sim$1500\,K,
and higher than the sublimation temperature of  silicate
grains (T$\sim$1000~K; \citealt{gd94}). 
Considering that our spatial resolution is limited to 
$\sim$380~pc, it is very likely that dust at higher temperatures 
exists closer to the central source, ruling out the possibility
of silicates as the main component of the nuclear dust grains.

Using the temperature of the blackbody as the average 
temperature of the graphite grains and 
a {\it K} band flux of 5.93$\times$10$^{-25}$ erg s$^{-1}$ 
cm$^{-2}$ Hz$^{-1}$ at 2.2\,$\mu$m found for the blackbody 
component after subtracting the underlying composite power-laws,
we can roughly estimate the dust mass associated with the 
bump. Following \citet{ba87}, the infrared 
spectral luminosity, in ergs s$^{-1}$ Hz$^{-1}$, of an 
individual graphite grain is L$_{\nu,\mathrm{ir}}^{\mathrm{gr}}=4\pi~a^{2}~\pi Q_{\nu} B_{\nu}(T_{\mathrm{gr}})$,
where $a$ is the grain radius, $Q_{\nu}= q_{\mathrm{ir}} \nu^{\gamma}$ is
the absorption efficiency of the grains and $B_{\nu}(T_{\mathrm{gr}})$ is the
Planck function for a grain of temperature $T_{\mathrm{gr}}$. Adopting,
as in \citet{ba87}, a value of $a$=0.05~$\mu$m for graphite grains
and $Q_{\nu}$=0.058 and setting $T_{\mathrm{gr}}$=1220~K, we find
$L_{\nu,\mathrm{ir}}^{\mathrm{gr}}$=9.29$\times$10$^{-18}$ 
ergs s$^{-1}$ Hz$^{-1}$.  

The total number of emitting grains (hot dust) can be approximated
as N$_{\mathrm HD} \approx L_{\mathrm NIR}/L_{\nu,\mathrm{ir}}^{\mathrm{gr}}$.

Finally, for graphite grains, with density $\rho_{\rm g}$=2.26 g~cm$^{-3}$,
M$_{\mathrm HD} \approx 4.12 a^{3} N_{\rm HD} \rho _{\mathrm g}$.
Taking Mrk~1239 at a distance of 79.7 Mpc  (z=0.00199, H$_o$=75~km\,s$^{-1}$/Mpc), we obtained 
$N_{\mathrm HD}$=4.51$\times 10^{46}$ and 
$M_{\mathrm HD}$=2.7$\times 10^{-2}$~M$\odot$. 

Table~\ref{masses} compares the mass of hot dust derived for
Mrk~1239 to that found in other AGNs. Our calculations show
that Mrk~1239 harbors the second largest mass of hot dust reported
to date in the literature for an AGN, only surpassed by NGC~7469. 
Note, however, that except for Mrk~766,
all other previous measurements are based on photometric
data, which do not take into account the underlying 
featureless continuum that we subtracted. Had we
used the peak flux of the continuum, we would have obtained
a value of 5.3~M$\odot$, ranking Mrk~1239 as the 
AGN with the largest content of hot dust found to date. 
The presence of dust near to sublimation temperature 
in Mrk~1239 has been largely predicted by dust emission models.
In fact, the onset of the broad strong bump of emission 
at 1$\mu$m, with peak at $\sim60\mu$m, is set by the
dust sublimation temperature of graphite grains at T$\sim$1500~K. 
Why in Mrk~1239 exists such a large amount of hot dust, to
the extent of creating a noticeable shoulder in the much broader
IR SED attributed to warm dust, cannot be tell from our data. 
No doubt that Mrk~1239 is an excellent target to be studied
spectroscopically with Spitzer in order to study the broad
band IR emission of this object.

\begin{table}
\begin{center}
\caption{Masses of hot dust found in AGNs} \label{masses}
\begin{tabular}{lcc}
\hline \hline
Galaxy  & Mass (M$\odot$)  &  Reference \\
\hline
Mrk~1239 & 2.7$\times10^{-2}$ & This work \\
NGC~7469 & 5.2$\times10^{-2}$ &  \citealt{ma98}    \\
Fairall~9 & 2.0$\times10^{-2}$ & \citealt{cwg89}     \\
NGC~3783 & 2.5$\times10^{-3}$ &  \citealt{gla92}    \\
Mrk~766  & 2.1$\times10^{-3}$ &  \citealt{ro05}   \\
NGC~1566 & 7.0$\times10^{-4}$ &  \citealt{bar92}   \\
\hline
\end{tabular}
\end{center}
\end{table}

Regarding to the location of the hot dust, our aperture size
implies that it must reside within the inner 380\,pc from
the centre. However, the high temperature of 
the dust imposes a tighter constraint to the location 
of its emitting region: the inner 100\,pc. 
This value is deduced by inspecting Fig.~3 of \citet{ma98}, 
where a plot of dust temperature as function of distance from
the nucleus was constructed for NGC~7469. 
Note that we have assumed that the dust in Mrk\,1239
has similar properties to that of NGC~7469.  
Further support to this distance can be obtained following 
the results for NGC~1068 \citep{ma00}. These authors concluded that
hot dust ($T_{\mathrm{gr}}$=1500~K) should be extremely
confined and located at a radius less than 4~pc. 
Although the precise location of the hot
dust cannot be distinguished from our
data, both scenarios are in accord to the polar 
scattering model of \citet{sm04}. Likely, the combined
effect of hot dust associated to the outer layers of the torus and the
inner hot dust of the polar scattering cone contribute
to enhance the bump observed in Mrk~1239. In fact, 10~$\mu$m
imaging of this object presented by \citet{gor04} shows 
clear evidence of bright extended emission in a
cone-like structure, with the apex of the cone located
at the nucleus. Although evidence of this extended
emission is not seen in our data, we especulate that
the hottest dust component should indeed contribute to the
observed bump.  No doubt, our results 
provides the first direct spectroscopic evidence of hot
dust in AGN and show the potential that NIR spectroscopy
has at unveiling that component.

If the strong NIR excess observed in Mrk~1239 is indeed
thermal emission from very hot dust, one can ask why we only see 
emission at a single temperature if one would
expect a range of temperatures? In that case, the NIR excess
should resemble more closely to a sum of blackbody curves of
decreasing temperatures. This question can be answered
if we remember that the IR emission in Mrk~1239 is largely dominated
by a much stronger bump, peaking at 60~$\mu$m (see, for
example, Figure~7 of \citealt{gru04}, where the broadband
continuum of this source is presented), indicating that a large
interval of dust temperatures indeed dominates the bulk of the 
IR emission over the hot one.

\section{Final Remarks} \label{final}

In this letter we have reported the first discovery of an
isolated NIR bump of emission in the NLS1 galaxy Mrk~1239. The continuum steeply rises
toward longer wavelengths redward of 1~$\mu$m, peaking at
2.2~$\mu$m, where it starts to fall smoothly with wavelength. This
excess of emission dominates the region between 1-5~$\mu$m. After
comparing the optical continuum with that of Ark~564, we found 
that the continuum in Mrk~1239 is reddened by E(B-V)=0.54, appreciably
larger for a type~1 object. This result agrees with polarimetry data,
which points out towards a dusty polar scattering region. In order to adequately reproduce
the NIR continuum, a Planck distribution of T$\sim$1200~K is needed to
account for the strong excess of emission over a featureless continuum
of power-law form. We interpreted this component in terms of very
hot dust, near its sublimation temperature, very likely 
located both in the upper layers of torus and close to the apex 
of the polar scattering region. 
If our hypothesis is correct, we have provided additional 
spectroscopic evidence of the presence of the putative torus of the 
unified model of AGNs.

\section*{Acknowledgments}

This research has been partly supported by the Brazilian agency
CNPq (309054/03-6) to ARA and the European 
Commission's ALFA-II program through its funding of the Latin-American 
European Network for Astrophysics and Cosmology, LENAC to XM. 
This research has 
made use of the NASA/IPAC Extragalactic Database (NED) which is 
operated by the Jet Propulsion Laboratory, California Institute of 
Technology, under contract with the National Aeronautics and Space 
Administration.

\label{lastpage}

\end{document}